
\magnification 1200


\def\reali{\hbox{\rm I\hskip-2pt\bf R}}
\def\complessi{\hbox{\rm I\hskip-5.9pt\bf C}}

\def\interi{\hbox{\rm{Z\hskip-8.2pt Z}}}
\def\naturali{\hbox{\rm I\hskip-5.5pt\bf N}}


\def\bsk{\bigskip}
\def\msk{\medskip}
\def\ssk{\smallskip}
\def\ni{\noindent}


\font\titfnt=cmbx10 at 14.40 truept


\newdimen\pagewidth \newdimen\pageheight \newdimen\ruleht
\hsize=31pc  \vsize=45pc  \maxdepth=2.2pt  \parindent=19pt
\pagewidth=\hsize \pageheight=\vsize \ruleht=.5pt
\abovedisplayskip=8pt minus 1pt
\belowdisplayskip=8pt minus 1pt
\abovedisplayshortskip=8pt minus 1pt
\belowdisplayshortskip=8pt minus 1pt

\baselineskip=14pt plus 1pt
\lineskip=15pt plus 1 pt

\mathsurround=1pt


\nopagenumbers
\def\testos{\null}
\def\testod{\null}
\headline={\if T\tpage{\gdef\tpage{F}{\hfil}}
            \else{\ifodd\pageno\rightheadline\else\leftheadline\fi}
           \fi}
\gdef\tpage{T}
\def\rightheadline{\hfil{\tensl\testod}\hfil{\tenrm\folio}}
\def\leftheadline{{\tenrm\folio}\hfil{\tensl\testos}\hfil}


\newcount\numref
\global\numref=1
\newwrite\fileref
\immediate\openout\fileref=ref.tmp
\immediate\write\fileref{\parindent 30pt}
\def\citaref#1{${[\the\numref]}$\immediate\write\fileref
              {\par\noexpand\item{{\the\numref . \enspace}}}\ignorespaces
              \immediate\write\fileref{{#1}}\ignorespaces
              \global\advance\numref by 1\ignorespaces}

\hfill  IFUP-TH 55/94
\bsk\bsk
\centerline{\titfnt Spectral stochastic processes}
\msk
\centerline{\titfnt arising in quantum mechanical models}
\msk
\centerline{\titfnt with a non--$L^2$ ground state}

\bsk\bsk
\centerline{J. L\"offelholz $ ^{*}$ }
\ssk
\centerline{\it Mathematisches Institut,
Universit\"at Leipzig, Germany}

\bsk
\centerline{G. Morchio}
\ssk
\centerline{\it Dipartimento di Fisica dell'Universit\`a and INFN,
           Pisa, Italy}

\bsk
\centerline{F. Strocchi}
\ssk
\centerline{\it Scuola Normale Superiore and INFN,
              Pisa, Italy}

\bsk\bsk\bsk
\ni {\bf Abstract}.  A functional integral representation is given for a
large class of quantum mechanical models with a non--$L^2$ ground state.
As a prototype the particle in a periodic potential is discussed:
a unique ground state is shown to exist as a state on the Weyl algebra,
and a functional measure ({\it spectral stochastic process\/})
is constructed on trajectories taking values
in the spectrum of the maximal abelian subalgebra of the Weyl algebra
isomorphic to the algebra of almost periodic functions.
The thermodynamical limit of the finite volume functional integrals
for such models is discussed, and the superselection sectors
associated to an \lq\lq observable\rq\rq\ subalgebra of the
Weyl algebra are described in terms of boundary conditions and/or
topological terms in the finite volume measures.

\bsk\bsk
\ni $ ^{*}$ supported by DFG, Nr. Al 374/1--2
\vfill
\eject

\def \A {{\cal A}}
\def \B {{\cal B}}

\def \Aq {{\cal A}_q}
\def \AE {{\cal A}_E}

\def \M {{\cal M}}
\def \D {{\cal D}}
\def \Dz {{\cal D}_0}

\def \vp {\varphi}
\def \H {{\cal H}}

\def \Hth {{{\cal H}_\theta}}
\def \Hz {{{\cal H}_0}}

\def \Ha {{{\cal H}_\alpha}}
\def \Ho {{{\cal H}_\omega}}

\def \Hpz {{{\cal H}_{\pi_0}}}

\def \Ua {U(\alpha)}
\def \Vb {V(\beta)}

\def \Oz {{\Omega_0}}
\def \Aobs {{\cal A}_{obs}}
\def \Aper {{\cal A}_{per}}
\def \psa {{\psi_\alpha}}
\def \pst {{\psi_\theta}}
\def \psz {{\psi_0}}

\def \piz {{\pi_0}}
\def \intinf {\int_{-\infty}^{+\infty}}

\def \dm {d\mu}

\def \emtH {\exp - t H}

\def \eamtH {e^{- t H}}

\def \Ldnz {L^2 (\Sigma , d\nu_0)}

\def \RMT {Riesz--Markov theorem}

\def \fjm {\varphi_{j,m}}

\def \sfjm { \{ \varphi_{j,m} \} }

\def \fjn {\varphi_{j,n}}

\def \dpjm {[0, 2 \pi m!/\alpha_j) }

\def \modjm {{\rm \ mod} \ 2 \pi m! /\alpha_j }

\def \modjmn {{\rm \ mod} \ 2 \pi \, (\min(m,n))! / \aj }
\def \aj {{\alpha_j}}
\def \fj {{\varphi_j}}

\def \Sj {\Sigma_j}
\def \S {\Sigma}

\def \SE {Q}
\def \pst {{\psi_\theta}}

\def \Tz {{T_z}}
\def \Tsz {{T^{*}_z}}
\def \qm {x_{-}}
\def \qp {x_{+}}

\def \ZpT {Z(\psi, T)}
\def \oth {{\omega_\theta}}

\def \dmT {{d\mu_T}}

\def \emW {e^{- \int_{-T}^T  W(x(s)) \, ds}}

\noindent
{\bf 1. Introduction}

\bigskip
The deep connection between (euclidean)
stochastic processes and Quantum Mechanics
(QM) relies on two basic properties: i) the representation
of the kernel of $\exp - \tau H$ as the integral of
a  conditional Wiener measure over trajectories
(Feynman--Kac formula), ii) the existence of the \lq\lq thermodynamical
limit\rq\rq\ ($T \nearrow \infty$) of the \lq\lq euclidean\rq\rq\
correlation functions
${< q(\tau_1) \, \ldots \, q(\tau_n) >}_{-T,T}$
\citaref{J. Glimm, A. Jaffe, {\it Quantum Physics, a functional
integral point of view\/}, second ed., Springer Verlag, 1987},
\citaref{B. Simon, {\it Functional Integration and Quantum Physics\/},
Academic Press, 1979}.

The latter property, which is in general not so much emphasized
as the first, is essential in order to construct a
{\it stationary measure\/} over trajectories, and plays an essential
r\^ole in the standard reconstruction of a QM Hilbert space,
with time translations described by unitary operators.
{}From a QM point of view, the existence of a stationary measure
is equivalent to the existence of a ground state for the
Hamiltonian $H$, an essential ingredient of the euclidean formulation
by analytic continuation   of the real time correlation functions,
and in fact a structural feature of Nelson's and
Osterwalder--Schrader's formulation of QM and Quantum Field Theory
(QFT) [1], [2], \citaref{G. Velo and A. S. Wightman eds.,
{\it Constructive Quantum Field Theory\/}, Erice School 1973,
Springer Verlag 1973}.

\ssk
Such property is satisfied by a large class of QM models (including
QFT with polynomial interactions), since the growth of the potential
at infinity implies the existence of a ground state in
$L^2(\reali^s , d^sq)$, and therefore of a stationary measure
over real trajectories
$q(\tau) \in \reali^s$, $\tau \in \reali$ the
euclidean time.

The above property however fails for an interesting class
of QM models, which include the free particle, particles in
periodic potentials and also QFT models (gauge QFT in the temporal
gauge are an example, due to the continuous spectrum of the Gauss
law operator). Such models can be more generally defined in
terms of representations of CCR Weyl algebras, and
then a ground state exists (and it is unique) as a non--regular
state over the Weyl algebra, $\A$, even if the Hamiltonian
has no point spectrum in the regular
(Schr\"odinger) representation (see Sect.3).

\medskip
For a functional integral representation of the correlation
functions of such ground states, one is then led
to choose a maximal abelian subalgebra $\A_c$ of $\A$
and to construct a measure on the trajectories $\sigma(\tau)$
taking values in the Gelfand spectrum  of $\A_c$;
the result is a measure on the product over (euclidean) times
of the spectrum of $\A_c$
({\it spectral functional measure\/}).

If $\A_c$ is the $C^{*}$algebra $\Aq$
generated by $\exp(i \alpha \cdot q)$, $\alpha \in \reali^s$,
then the functional measure is defined on trajectories
in the spectrum of
the algebra of almost periodic functions, to which
$\Aq $ is isomorphic;
the variables $q_i(\tau)$ are well defined on the support
of the measure, and describe the spectrum of QM operators,
{\it only} when the representation of $\Aq$  is {\it regular}, i.e.
$\exp(i \alpha \cdot q)$ is strongly continuous in $\alpha$.

\medskip
{}From a general functional--measure theoretical point of view,
the construction of measures is usually done
on the basis of Minlos' theorem [2], \citaref{T. Hida,
{\it Brownian motion\/}, Springer Verlag 1980},
starting from (positive) functionals, i.e. expectations of
$\exp(i \alpha \Phi(f)) \, , \; f \in {\cal S}$,
continuous in $\alpha$;
when such  continuity property fails, the above functionals
cannot be represented by measures on
the variable $\Phi$ in a space of distributions.
Examples, which occurs in the models discussed below, are those
of functionals defined by
an ergodic mean (\lq\lq flat distribution\rq\rq ) in some real variable.
Quite generally, however,
the Gelfand and Riesz--Markov theorems imply
that such \lq\lq non--regular\rq\rq\
functionals can be represented by measures
defined {\it on the spectrum of the algebra generated by the
above exponentials\/}, and $\Phi(f)$ can then be interpreted as a
{\it \lq\lq non--regular random variable\rq\rq \/}.

\medskip
The above problems have strong implications
on the infinite \lq\lq volume\rq\rq\  limit ($T \nearrow \infty$) of
stochastic processes constructed
with some fixed \lq\lq end point\rq\rq\ $x(T) = \bar x$;
already the brownian motion (free particle
in the quantum mechanical interpretation) is an example,
but problems become acute when the control on the model
is not explicit.
In fact, for QM models,
the functional measures corresponding to different
Hamiltonians have disjoint support in an infinite time interval, and
the standard strategy for their construction is to work in
finite \lq\lq volume\rq\rq ,  $-T \leq \tau \leq T$,
give boundary conditions in time, and then take
the limit $T \nearrow \infty$.
In the \lq\lq non--regular\rq\rq\ case,
the finite volume measures do not converge
as  measures  on real trajectories $x(\tau)$;
they do however converge
(see Sect.4) as measures on $\SE \equiv \prod_{\tau} \S$,
the product over all times of the spectrum $\S$ of the
algebra of almost periodic functions.

In the theory of stochastic
processes similar considerations and results apply,
if one looks for the existence of stationary measures:
they must in general be defined on
the spectrum of a suitable algebra of
bounded functions of the trajectories.

\medskip
The aim of this letter is:

\noindent
i) to provide a general mathematical framework for
the construction of functional measures defined on trajectories
$\sigma (\tau)$ taking values
in the spectrum of the algebra of almost periodic functions;

\noindent
ii) to provide a functional integral representation of the
non--regular states over the Weyl algebra
arising as ground states of Hamiltonians with periodic potentials;

\noindent
iii) to discuss the thermodinamical limit ($T \nearrow \infty$)
of the finite volume functional measures for such models;

\noindent
iv) to describe the superselection sectors which arise associated
to an \lq\lq observable\rq\rq\ subalgebra (of the Weyl algebra),
and their construction in the thermodynamical limit as a result
of boundary conditions and/or
\lq\lq boundary terms\rq\rq\ (corresponding to {\it topological
terms} in the action) in the finite volume functional measure.

\bigskip\bigskip\goodbreak\noindent
{\bf 2. Spectral stochastic processes}.

\bigskip
The basic structure for the algebraic formulation of QM
is the CCR algebra generated by the Weyl
operators $U(\alpha)=\exp i q \cdot \alpha   $,
$V(\beta) = \exp i p \cdot \beta  $, $\alpha , \beta \in \reali^s$,
and defined by the Weyl relations
$$ U(\alpha) V(\beta) =
     V(\beta) U(\alpha) \exp i \alpha \cdot \beta \ \ . \eqno(2.1) $$
There is a unique $C^{*}$ norm on the CCR algebra
\citaref{J. Manuceau, M. Sirigue, D. Testard and A. Verbeure,
Comm. Math. Phys., {\bf 32}, 231, (1973)},
the operator norm in the  Schr\"odinger
representation; by $\A$ we will denote the $C^{*}$ algebra obtained
by norm completion.

States are defined as positive linear functionals on $\A$.
Each state $\omega$ defines, by the usual
Gelfand--Naimark--Segal (GNS) construction,
a representation of $\A$ in a Hilbert space $\Ho$. A representation
is called regular if the unitary groups $\Ua \, , \; \Vb$ are
(strongly) continuous in their parameters. We are interested
in representations in which the group $\Ua$ is not continuous.
For simplicity we put $s=1$ in the following.

The (maximal commutative) $C^{*}$subalgebra $\Aq$
generated by $\Ua = \exp i \alpha q $, $\alpha \in \reali $,
is the algebra of almost periodic functions $f(x)$, $x$ a
real variable
\citaref{See, e.g., C. Berg, Introduction to Almost
Periodic Functions of Bohr, in Math.--Fys. Medd, {\bf 42:3},
(1989) and references therein. See also N. I. Akhiezer and
I. M. Glazman, {\it Theory of Linear Operators in Hilbert
Space\/}, vol.1, third ed., Pitman 1981};
by the Gelfand isomorphism, it coincides with the algebra of
continuous functions on a compact space, $\S$, which we will
characterize below.

\medskip
To construct functional integral representations of
a state $\omega$, without regularity assumptions,
given a time evolution defined by a Hamiltonian (bounded below)
in $\Ho$, we start from the expectation of products of elements of
the abelian subalgebra $\Aq$, at ordered imaginary times;
we write such expectations as
$$ \omega(A_1(t_1)) \ldots A_n(t_n))   \ \ \ . $$
The algebra $\Aq$  plays here the r\^ole usually taken by the algebra
of bounded (continuous) functions of a real variable $q$, which
is represented in $\Ho$ only under regularity assumptions
(strong continuity of $\Ua$).

If such expectations define a {\it positive} linear functional
on the (\lq\lq euclidean\rq\rq ) abelian algebra
$\AE (T_1 , T_2) $
generated, through linear combinations and closure in the Sup norm,
by the functions
$$\exp i \sum_i \alpha_j q(t_j) \ \  , \ \ \
q(t_j) \in \reali \ \ , \ \ \  a_j \in \reali \ \ ,
\ \ \ t_j \in [T_1 , T_2] \ \ ,  \eqno(2.2)$$
then the Riesz--Markov theorem represents such a functional
in terms of a Baire measure,
with a unique regular Borel extension, on the
spectrum $\SE$ of $\AE$ (here $[T_1, T_2]$ may be  a finite or
infinite interval); see, e.g.,
\citaref{M. Reed, B. Simon, {\it Methods of Modern Mathematical
Physics\/}, vol. 1, Academic Press, Sect. 4.4}.
It is easy to see that $\SE $ is the product over
$t \in \reali$ of copies of the spectrum $\S$ of $\Aq$.
It can therefore be identified with
 a space  of {\it spectral trajectories}
$\sigma(t) $, i.e. functions of the real variable
$t$ taking values in the spectrum $\S$ of $\Aq$.

The above arguments show that in general,
without regularity assumptions,
{\it Nelson positivity} of
the euclidean correlation functions of $\Aq$ implies
their representation as integrals over {\it spectral} trajectories:
if $f_i(\sigma)$ is the Gelfand image of the element
$A_i$ of $\Aq$, then
$$ \omega(A_1(t_1)) \ldots A_n(t_n)) =
\int \dm (\sigma (t))  f_1(\sigma(t_1)) \ldots
           f_n(\sigma(t_n))     \eqno(2.3)    $$
The above considerations on the construction of measures
on trajectories in the spectrum of the algebra of quasi--periodic
functions are largely independent from their motivations in
Quantum Mechanics and apply to a (a priori) much larger class of
problems like, e.g., the construction of a stationary measure
for the brownian motion.

\medskip
To characterize  measures  on spectral trajectories
it is convenient to have an explicit construction of the spectrum
$\S$ of $\Aq$.
We recall that the Gelfand spectrum of a commutative $C^{*}$ algebra
$\B$ (with identity) is the space $\M$ of
multiplicative linear functionals $M : \B \to \complessi \, $,
with the weak topology defined by $\B$ on $\M$.
By Gelfand's theorem, $\B$ is isomorphic to the
algebra $C(\M)$ of the continuous functions over its spectrum,
with the Sup norm \citaref{M. Naimark, {\it Normed Rings\/},
Nordhoff, 1964; M. Takesaki, {\it Theory of Operator Algebras I\/},
Springer Verlag 1979}.
The  proof of the following results is given in
ref. \citaref{J. L\"offelholz, G. Morchio, F. Strocchi,
in preparation}.

\medskip\goodbreak\noindent
{\bf Proposition 1}.
{\it
The Gelfand spectrum $\Sigma$ of
$\Aq$ has the following representation:
given a Hamel basis $ \{ \alpha_j \}$,
for the reals over the rationals, a point $\sigma \in \Sigma$
is a generalized sequence of real numbers
$$ \fjm \in \dpjm \ ,\ \ \fjm = \fjn \ \ \modjmn \ , \
\ n,m \in \naturali $$
The corresponding functional $M_\sigma$ is given by
$$M_\sigma (U(\aj /m!)) =
\exp (i \aj \fjm / m!) \equiv
\exp [{i \over m!} ( \aj \fj + 2 \pi
   \sum_{l=1}^m \, N_{j,l} (l-1)!)]
\eqno(2.4) $$
with $\fj \in [0, 2 \pi / \aj )$, and the integers
$ N_{j,l} \in [0, l-1]$.

\ni
$\Sigma$ is the topological product of the spaces $\Sigma_j $,
consisting of the above sequences with $j$ fixed.
The topology is defined on $\Sigma_j$ by the basis of neighbourhoods
$$ I_{n,\epsilon} (\{ \fjm^0 \}) =
\{ \{ \fjm \} : \, | \fjn - \fjn^0 |  < \epsilon   \ \}  \
 \ \  . \eqno(2.5)  $$

\ni
$\Sigma $ and all the spaces $\Sigma_j$ are compact.

\ni
The points $\sigma$ of $\S$ such that $M_\sigma(\Ua)$
is continuous in $\alpha$, and therefore of the form
$M_\sigma(\Ua) = \exp i \alpha x$, $x \in \reali$,
define a dense set of \lq\lq real points\rq\rq\ of $\Sigma$,
of the form
$$ \fjm = x \ \ \modjm \ \ , \ \ \ x \in \reali \ \ . \eqno(2.6) $$
We will write $x(\sigma)$ for the (unique) real number $x$
corresponding, via eq.(2.6), to a \lq\lq real point\rq\rq\
$\sigma \in \S$.

\noindent
The translation automorphisms
$\Tz  :  \exp i \alpha q \mapsto
\exp i \alpha (q+z) $ define on $\Sigma$, by duality, a
continuous
one--parameter group $\Tsz$, $z \in \reali $:
$$\Tsz \sfjm = \{ \fjm + z \ \ \modjm  \} \ \ . \eqno(2.7) $$
$\Tsz \sigma$ will also be denoted by $\sigma + z$}.

\medskip\noindent
The use of the Hamel basis \citaref{G. Hamel, Math. Annalen,
{\bf 60}, 459 (1905)}
is necessary for the spectrum of the
algebra of all periodic functions. One may however consider
separable subalgebras, e.g., the algebras generated by exponentials
with periods which are {\it rational} combinations of a finite
numbers of periods, $L_i$; Proposition 1 then applies with the
Hamel basis replaced by $\{ \alpha_i = 1/L_i \}$.
Most of the phenomena and results can also be
discussed on the basis of such algebras; their spectra
are finite products of spaces $\Sj$, which have a
much richer structure then the spectra
of the subalgebras defined by {\it integer} combinations of $\alpha_i$
(which are tori).

\medskip
By Proposition 1, the Borel sets of $\S$ are generated
(by countable union and complements)
by the open sets in a finite number of variables $\fjm$.
Regular Borel measures on $\S$ are therefore identified by their
restriction to tori of the form
$$  \{ \fjm \in \dpjm \ \ , \ \ \ (j,m) = (j_1,m_1) \ldots
  (j_k,m_k) \} \ \ . \eqno(2.8)   $$

It is not difficult to show [9] that the
set of \lq\lq real points\rq\rq\ of $\S$ is a Borel set in $\S$, and
that the Borel structure defined on the \lq\lq real points\rq\rq\
by the spectral topology
coincides with the standard Borel structure of the real line.
Therefore, every Borel measure on the  real line defines a
Borel measure on $\Sigma$ with \lq\lq real\rq\rq\ support.
There are, however, many (regular) Borel measures on
$\S$ with support disjoint from the real points.
The following ones are of special interest:
given any positive almost periodic function $M(x)$, the ergodic
mean
$$ \lim_{L \to \infty} 1/2L \; \int_{-L}^L M(x) f(x) dx  \eqno(2.9) $$
defines a positive functional on $f \in \Aq $,
and therefore a measure on $\S$;
for $M(x) = 1$ this measure,
denoted by $d\nu_0$,
is given
by the usual (normalized) Lebesgue measure on the tori (2.8).
The support of all such measures is disjoint from the set of
the \lq\lq real points\rq\rq\ of $\S$ [9].

\medskip
A class of measures on
$\SE = \prod_{t \in \reali} \Sigma$, corresponding to
\lq\lq locally real stochastic processes\rq\rq\ on $\S$
 can be constructed as follows:
given any stochastic process defined on the real line
and starting at $z(0) = 0$,
identified by the measures
$d\rho (z(t_1), \ldots , z(t_n))$,
and any Borel measure $d\nu (\sigma)$ on $\S$, a measure on
$\SE $ is defined by
the measures
$$ d\mu (\sigma(0), \sigma(0) + z(t_1), \ldots , \sigma(0) + z(t_n))
\equiv  d\nu (\sigma(0)) \;
   d\rho (z(t_1), \ldots , z(t_n))    \ \ \ .  \eqno(2.10)  $$
Such a measure is supported  on trajectories
of the form $ \sigma(t) = \sigma(0) + z(t) $, with
spectral translations $z(t)$ in the support of $d\rho$.

For example, if $d\rho$ defines the brownian motion
starting at $z(0) = 0$, and $d\nu = d\nu_0$,
then $d\mu (\{ \sigma(t) \})$ defines a stationary Markov process
on $\S$ (an explicit control follows from the
above characterization of $d\nu_0$ on all
tori (2.8)).
This construction formalizes the fact that
\lq\lq the ergodic mean on the real line is invariant
under the brownian motion\rq\rq.

In the following section we show that
measures of the form (2.10), with
$d\nu = d\nu_0$, represent the ground state
of any quantum mechanical Hamiltonian of a particle in a
periodic potential.

\bigskip\bigskip\noindent
{\bf 3. Ground states for periodic potential and their
functional integral representation}.

\bigskip
Existence of the Hamiltonian and of the ground state for
particles in periodic potentials is controlled by
the following results [9]:

\medskip\goodbreak\noindent
{\bf Proposition 2.}
{\it
i) Given any
bounded, measurable and periodic potential $W(x)$ (of period $1$),
there exists one and only one irreducible representation
of the Weyl algebra in which the Hamiltonian
$$ H =  p^2/2 + W  $$
is well defined as a strong limit of elements
of $\A$ on a dense domain, and has a ground state.

\noindent
ii) Such a representation is independent of $W$ and is the
unique (non--regular) representation $\piz$ in which
the subgroup $V(\beta) , \, \beta \in \reali$ is regularly represented and its
generator
$p$ has a {\it discrete spectrum}.

\noindent
iii) The Hilbert space $\H$ of $\piz$ is given by the
Gelfang--Naimark--Segal (GNS) construction over the state $\Oz$
on $\Aq$ defined by the ergodic mean in $q$;
its vectors are formal sums
$$ \psi(x) = \sum_{n\in \interi} a_n \exp i \alpha_n x \ \  , \ \ \ \
x \in \reali \ , \ \
\{a_n\} \in l^2 \ \ ,                             \eqno(3.1) $$
and the Weyl operators are represented as
$$U(\alpha) \psi(x) = \exp (i \alpha x) \; \psi(x) \ \ , \ \ \
  V(\beta)  \psi(x) = \psi(x+\beta)   \ \ .        \eqno(3.2)$$
The scalar product in $\H$ is given by }
$$ (\psi,\psi) = \sum_n |a_n|^2 =
{\rm ergodic \ mean \ } (|\psi|^2) \ \ \ .  \eqno(3.3) $$

\msk
It follows that $\H$ contains a dense subspace described by wave
{\it functions} $\psi(x)$.
Moreover, $\H$ is the direct {\it sum}, over $\alpha \in [0,2\pi]$,
 of spaces $\Ha$ of wave functions of the form
$\exp (i \alpha x) f(x)$,
with $f(x)$ locally $L^2$ and $f(x+1) = f(x)$.
In $\Ha$ the operator $p$ is
represented by $ - i d/dx \, $, with boundary conditions
$\psi(1) = \exp(i \alpha) \, \psi(0)$.

The above {\it direct sum} over $\alpha$
corresponds to the reduction of $\piz$ into
irreducible representations of the
algebra $\Aper \subset \A$, generated by
$ \{ \exp i (\beta p + nq) $,
$ \beta \in \reali  , \, n \in \interi \, \}$.
The Hamiltonian $H$ is affiliated to the strong closure of
$\Aper$ in $\piz$ and leaves the spaces $\Ha$ invariant.
We denote by $\Dz$ the subspace of the vectors of the form
$A \cdot 1$, $A \in \Aq$, and by
$\D$ the linear span of the spaces $\Ha$.

\medskip
The following Proposition extends the ordinary
path integral representation of the
kernel of $\exp - t H $
to the non regular representation $\piz$;
uniqueness of the ground state in $\H$ then follows
by a Perron--Frobenius argument.

\medskip\goodbreak\noindent
{\bf Proposition 3}.[9]
{\it
For all $\psi \in \D$,
$$ ( e^{- t H} \  \psi)(x) = \intinf dy \, K_t(x,y) \psi(y) \ \,
  \eqno(3.4) $$
where $K_t(x,y)$ is the kernel of $\exp - t H$ in the
Schr\"odinger representation, which is strictly positive
by the usual Feynman--Kac representation.

\noindent
The ground state $\psz $ is unique (up to a phase);
it belongs to $\Dz \cap \Hz$
and its wave function can be chosen to be positive;
it is cyclic for the algebra $\Aq$.
In each  $\Ha$,
$H$ has a unique lowest energy state $\psa$
($H \psa = E(\alpha) \psa$).
         }

\medskip\noindent
Positivity of $\psz (x)$ and of the kernel of
$\exp - t H$ in $\piz$ imply Nelson positivity
for the ordered imaginary time correlation
functions of $\Aq$ on $\psz$. In fact,
normalizing to zero the ground state energy $E(0)$, they are given by
$$ \omega (e^{i \alpha_1 q(t_1)} \,  \ldots
\,  e^{ i \alpha_n q(t_n)} ) \equiv $$
$$ \equiv (\psz \, , \, e^{i \alpha_1 q} e^{-H (t_2 - t_1)}
    e^{i \alpha_2 q} \ldots e^{-H (t_n - t_{n-1})}
    e^{i \alpha_n q} \psz )   \ \ .  \eqno(3.5) $$

By the group property of
$\exp - t H$ and the normalization of the
ground state energy, eq.(3.5) defines a functional on $\AE$.
The representation of the scalar product
in $\H$, eq.(3.3), gives the r.h.s. of eq.(3.5)  the form
$$  {\rm ergodic \ mean \ } (x_1)
     \intinf dx_2 \, \ldots \, dx_n \;
     \psz (x_1) \,   e^{i \alpha_1 x_1} \, $$
$$      K_{t_2 - t_1} (x_1,x_2) \,
      e^{i \alpha_2 x_2} \, \ldots \,
     K_{t_n - t_{n-1}} (x_{n-1},x_n) \,
    e^{i \alpha_n x_n} \psz (x_n) )  \ \ ,   \eqno(3.6) $$

The functional $\omega$
is therefore positive, and by the \RMT  \ it defines
a measure $d\mu $
on $\SE$, i.e. on
{\it spectral trajectories} $t \to \sigma(t) \in \S $.

\medskip
The measure $d\mu$ can be written explicitely,
making use of the measure $d\nu_0$ on $\S$
corresponding to the ergodic mean, eq.(2.9) with $M=1$.
In fact, from Proposition~2 it  follows that
$\H$ can be represented as  $\Ldnz$, and Proposition~3
implies that $\psz$ is represented by a continuous function
$F_0(\sigma)$, which is determined by its values
$\psz(x)$  on the \lq\lq real points\rq\rq\ $x \in \S$.

Moreover, from Proposition 3 it follows that
for all continuous functions $F$ on $\Sigma$ and all
real points $\sigma$,
$$ ( \eamtH \, F )(\sigma) = \intinf dz \; K_t (x(\sigma), x(\sigma + z))
    \; F(\sigma + z)    \eqno(3.7) $$
where we have used the translation group
on $\Sigma$, eq.(2.7).

Eq.(3.7) extends to all points $\sigma \in \Sigma$, since the
l.h.s. is continuous ($\emtH$ maps $\Dz$ into $\Dz$) and
the resulting kernel $ K_t (\sigma, \sigma + z)$ is continuous in $\sigma$
and $L^1$--bounded in $z$, uniformly in $\sigma$.
Furthermore, the Feynman--Kac representation of $K_t(x,y)$,
implies, for real points $\sigma$ in $\Sigma$,
$$  K_\tau(\sigma, \sigma + z) = \int d\rho_{[0,\tau],0,z} (z(t)) \;
\exp ( - \int_0^\tau W(\sigma + z(s)) \, ds )  \ \ \ , \eqno(3.8)$$
where $d\rho_{[0,\tau].0,z} (z(t))$ is the conditional Wiener
measure on paths $z(t)$, $0 \leq t \leq \tau$, with $z(0) = 0$,
$z(\tau) = z$.
The extension of eq.(3.8) to all points of $\Sigma$
is immediate for continuous potentials, to which we will
now restrict for simplicity.

By inserting eq.(3.8) in eq.(3.4), we obtain a (generalized)
Feynman--Kac representation of the propagator in $\H$, in terms of
a functional measure on the spectrum of $\AE$, i.e. on
trajectories $\sigma(t)$ with values in $\Sigma$.
Eq.(3.6) now gives the following

\medskip\goodbreak\noindent
{\bf Theorem 4}.
{\it
The unique ground state $\omega$ of a particle in a
continuous periodic potential has the representation
$$ \omega (e^{i \alpha_1 q(t_1)} \,  \ldots
\,  e^{ i \alpha_n q(t_n)} ) = $$
$$  = \int d\nu_0 (\sigma) \, F_0 (\sigma)
      \, \int d\rho_{[t_1, t_n],0} (z(t)) \;
 e^{ - \int_{t_1}^{t_n} W(\sigma + z(s)) \, ds }  $$
$$   e^{i \alpha_1 q}(\sigma) \,
e^{i \alpha_2 q}(\sigma + z(t_2)) \, \ldots \,
e^{i \alpha_n q}(\sigma + z(t_n)) \,
  F_0 (\sigma + z(t_n))  \ \ \ ,   \eqno(3.9) $$
where
$ d\rho_{[t_1, t_n],0} (z(t))$ is the Wiener measure on paths
$z(t)$, $t_1 \leq  t \leq t_n $, starting at
$0$ at $t_1$,  $F_0$  the representation of the ground state
in $L^2 (d\nu_0)$, and $e^{ i \alpha q} (\sigma)$ the
Gelfand image of $e^{i \alpha q}$.

\noindent
The r.h.s. of eq.(3.9) defines a (Borel) measure $d\mu$
on $\SE$ with support on trajectories of the form
$ \sigma (t) = \sigma (0) + z(t , \sigma(0)) $,
with $ z(t , \sigma(0)) $ Brownian  trajectories in $\reali$.
The restrictions of $d\mu$ to bounded time intervals
$[-T,T]$ are given by
$$  d\mu_{[-T,T]} (\sigma + z(t)) =  $$
$$  = d\nu_0 (\sigma) \; F_0 (\sigma)
      \;  d\rho_{[-T,T],0} (z(t)) \;
 e^{ - \int_{-T}^{T} W(\sigma + z(s)) \, ds } \;
  F_0 (\sigma + z(T))  \ \ \ .   \eqno(3.10) $$

\noindent
By using the boundedness of
$F_0 (\sigma)^{-1}$, $\dm$ can be put in the form (2.10).

\noindent
$\dm$ is  the weak limit, for $T \nearrow \infty$ of
spectral functional measures
$  d\mu_{[-T,T]}^f (\sigma + z(t))   $
in finite time intervals, defined by
$$   Z(f,T)^{-1} \;  d\nu_0 (\sigma) \,  f(\sigma) \;
    d\rho_{[-T, T],0} (z(t)) \;
 e^{ - \int_{-T}^{T} W(\sigma + z(s)) \, ds } \;
  f (\sigma + z(T))     \eqno(3.11) $$
for any non--negative $f(\sigma) \in \Ldnz$, in particular $f = 1$.
$Z(f,T) $  is given by $(f , \, \exp -2TH \, f) $ and converges to
$ |\int d\nu_0 \, F_0(\sigma) f(\sigma) |^2 $ as
$T \nearrow \infty$. }

\medskip
\noindent
{\bf Proof}. The last point follows from the fact that the
integral of the generators of $\AE$ with the measure (3.11)
has the quantum mechanical interpretation
$$ (\psi_f \, , \, e^{-H(t_1 + T)} \, e^{i \alpha_1 q}
, \, e^{-H(t_2 - t_1)} \, \ldots \,
 e^{i \alpha_n q}  \, e^{-H(T - t_n)} \,  \psi_f) \ \ \ , $$
which converge, for $T \nearrow \infty$, to
$ |(\psi_f , \psz)|^2  \
   \omega (e^{i \alpha_1 q(t_1)} \,  \ldots
   \,  e^{ i \alpha_n q(t_n)} ) $,
as a consequence of the spectral properties of the Hamiltonian in
$\H$ [9].
Since $\psz(x)$ is positive, one can take any $f(\sigma) \geq 0 $.

\bsk\bsk\goodbreak
\noindent
{\bf 4.  Infinite volume limit of measures over real trajectories}

\bigskip
{}From a constructive point of view it is important to
recover the above spectral functional measure in the limit
$T \nearrow \infty$ starting from standard functional measures
over real trajectories in bounded time intervals $[-T,T]$.
As we shall see, it is crucial to realize that such measures
do not converge as measures on real trajectories, but only as measures
on $\SE$ (measures on real paths uniquely define measures
on $\SE$ with support on \lq\lq real trajectories\rq\rq\ in $\SE$,
see Sect.2).

Given any Hamiltonian of a particle in a bounded measurable
periodic potential, if the times are restricted
to a bounded interval $[-T,T]$,
the expectation of the euclidean variables $f(q(t))$,
on any state $\psi \in L^2$ have the usual representation
$$ ( e^{-H(T + t_1)} \, \psi \, , \, f_1 \,
     e^{-H(t_2 - t_1)} \, f_2  \ldots
     e^{-H(t_n - t_{n-1})} \, f_n
\, e^{-H(T - t_n)} \, \psi ) \big /  || e^{-HT} \psi ||^2  =  $$
$$ = \ZpT^{-1}  \int  d\qm \, d\qp \,
\overline \psi(x_{-}) \, \psi(x_{+}) \;
   \int d\mu_T (x(t)) \;
   f_1(x(t_1)) \ldots
    f_n(x(t_n))    \eqno(4.1) $$
with
$$ d\mu_T (x(t)) \equiv
d\rho_{[-T,T],x_{-},x_{+}}  (x(t)) \  \emW    $$
and  $ \ZpT $ the usual normalization factor.

In the limit $T \nearrow \infty$ the correlation functions (4.1) of
the variable $f(x)=x$ are in general divergent; in fact, as it also
follows from the results given below,
the expectations of all bounded functions of
$x$ vanishing at infinity converge to zero in the infinite
time limit, so that, e.g.,
the expectation of $x(0)^2$ has a divergent lower bound as $T$
goes to infinity.
The point is that brownian trajectories in a bounded potential
wander over larger and larger regions for large times.
A more appropriate  choice of variables is therefore essential,
and in fact the expectations $(4.1)$ have a limit
in the formulation based on the Weyl algebra, i.e.,
for $f \in \Aq$.

\bigskip\goodbreak\noindent
{\bf Theorem 5}.
{\it The l.h.s. of
eq.(4.1), with $f_i$ in $\Aq$ and
$\psi(x) $ in $L^1 \cap L^2$ such that
$ \int \psz(x) \psi(x) \, dx \neq 0 $
converges for $T \nearrow \infty$ to the correlation functions,
eq.(3.5), of $\Aq$ on
the unique ground state $\psz$.
The measures defined by eq.(4.1) converge therefore weakly
on $\AE$ to the measure $d\mu (\sigma(t))$
constructed in Sect.3, eqs.(3.9),(3.10).
   }

\medskip\goodbreak\ni
{\bf Proof}.
The proof follows [9] from the integral decomposition of $L^2$
in spaces $\Ha$ (see Sect.3), the spectral analysis of
the Hamiltonian, the representation of $\Hpz$ given in Sect.3,
the uniqueness of the ground state
and some estimates in $T$ of scalar products and normalization factors.

\medskip\goodbreak
The measures $\dmT $ do not therefore converge as
measures over trajectories $ x(t) $,
but only as (\lq\lq locally brownian\rq\rq ) measures on
spectral trajectories  $\sigma(t)$.
Also when restricted to the $\sigma$--algebra generated by the
variables at a fixed time, e.g. $t=0$, the limiting measure
has support disjoint from the support of the finite time measures,
since the latter, eq.(4.1), are supported, as measures on $\S$, on its
\lq\lq real points\rq\rq.
{}From the characterization, eq.(3.10), of the support of
$d\mu$, it also follows that the problems of convergence
of the measures (4.1) can be  \lq\lq reduced to one time\rq\rq;
i.e., they arise in the construction of a measure on $\S$,
the spectrum of the algebra $\Aq$ at one time, invariant under
an evolution defined by (perturbed) brownian translations of $\S$.
In quantum mechanical terms, the problem is the ground state
(a non--regular state on $\Aq$), {\it not} the Feynman--Kac
representation of $\exp - t H$.

The disjointness of the measure defined by the vacuum
on the spectrum of the algebras generated by the variables
at fixed time, more generally in
{\it bounded\/} time intervals, which is typically
related to ultraviolet problems, occurs here for
{\it infrared\/} reasons, and one may speak of an
{\it  infrared renormalization}.
It corresponds to
a change of representation for the {\it local} algebras,
and has nothing to do with
the usual fact that in infinite volume (or time) the
functional measures with different interactions are disjoint;
the latter property is based on ergodicity in time and is
a functional measure version of Haag's theorem.

\bsk\bsk\goodbreak\ni
{\bf 5. Superselection sectors, boundary terms, winding numbers}.

\msk
The construction of a functional integral representation of
non regular states over
the Weyl algebra allows for the discussion of models where the Weyl
algebra plays the r\^ole of a {\it field\/} algebra, and a subalgebra
is identified with the {\it observable} algebra $\Aobs$.

We consider the case where $\Aobs = \Aper$.
As mentioned in Sect.2, $\Hpz$ decomposes into irreducible
representations of
$\Aper$, labelled by an angle $\theta \in [0, 2 \pi) $
and defined by the \lq\lq ground states\rq\rq\ $\oth$,
$\oth (A) = (\pst , A \pst )$, $A \in \Aper$.

{}From eqs.(3.5)--(3.10) a functional representation of $\oth$ on
trajectories in the circle
(the spectrum of the maximal abelian subalgebra
$\A_{q,per}$ of $\Aper$ generated by
$\{ \exp i 2 \pi n q \, \; n \in \interi \}$) follows
by taking as boundary condition the wave function $\psi_\theta$
of the unique ground state in $\Hth$.
In fact, $\psi_\theta(\sigma)$ is of the form
$ (\exp i \theta q)(\sigma) \, R_\theta(\sigma) $ with
$R_\theta$  a function depending only on one
\lq\lq spectral angle\rq\rq\ $\vp \in [0,1) = S^1 / 2\pi$;
the factors $\exp i \theta q$ can be inserted in the kernels
$K_t$, where they become functions of $z$;
since the Hamiltonian commutes with periodic translations, the kernel
$ K_t$ is a function of
$\vp$ and $z$; $d\nu_0$
becomes $d\vp $ and
$ \oth ( e^{i 2 \pi n_1 q(t_1)} \,  \ldots
\,  e^{ i 2 \pi n_k q(t_k)} )    $
has the integralrepresentation
$$   \int \prod_{i=1}^k  d\vp_i  \
R_\theta (\vp_1)   \;
{\cal  L}^\theta_{t_2 - t_1} (\vp_1 , \vp_2)  \;
 \ldots \,
{\cal L}^\theta_{t_k - t_{k-1}} (\vp_{k-1},\vp_k) \;
\prod_{j=1}^k e^{i 2 \pi  n_j \vp_j} \;
  R_\theta (\vp_k)    \eqno(5.1) $$
where
$$ {\cal  L}^\theta_t (\vp_1 , \vp_2)  \,  \equiv  \,
   \exp E_\theta t \,
   \sum_{n= - \infty}^\infty
   K_t (\vp_1 , \vp_2 +  n) \,
   \exp ( i \theta (\vp_2 - \vp_1 + n)) \ \ \  .   \eqno(5.2) $$

{\bf Proposition 6}.
{\it Eq.(5.1) defines a
bounded complex measure $d \gamma^\theta_T $
on trajectories $ t \in [-T,T] \to \vp (t) \in S^1 / 2\pi$,
with $\int d \gamma^\theta_T = 1$.
In the case $\theta \neq 0$ its
total variation, which is bounded by $\exp 2 T E(\theta)$,
diverges exponentially for $T \nearrow \infty$[9].
Moreover, Nelson positivity does not hold for the
correlation functions defined by $\oth$, $\theta \neq 0$,
even for variables in $\Aper$.  }

The representation (5.1) is given e.g. in
\citaref{L. S. Schulman, {\it Techniques and Applications of
Path Integration\/}, J. Wiley 1981},   where
$\theta$ corresponds to the apparently \lq\lq irrelevant\rq\rq\
addition of a time derivative $i \theta \, d\vp(t) /dt$
to the Lagrangean.
The point is that
the variable
$\exp i \theta  \vp$ belongs to $\Aq$, but not to
$\Aper$, and therefore
the \lq\lq topological\rq\rq\ term
$\exp i \theta \int_{-T}^T d\vp(t) /dt$
is equivalent to a change of boundary conditions
only in the formulation based on $\Aq$, i.e.
for functional integrals based on the spectrum of $\Aq$.
On the contrary, in the formulation based on trajectories
in the circle, (the spectrum of $\A_{q,per}$)
the r\^ole of $\theta$ is that of an additional term in the
Hamiltonian (see the definition of
$ {\cal  L}^\theta_t $, eq.(5.2)), and there is no relation
with boundary conditions.

\medskip
Such relation can be made explicit also in terms
of infinite time limit
of functional measures over real trajectories $\vp(t)$.
For boundary conditions $\psi \in L^1 \cap L^2$,
one obtains [9] the correlation functions
on the lowest energy state(s) of $\Hpz$
which appear in the integral decomposition of $\psi$
into components $\psi_\theta \in \Hth$
In particular, the result is the ground state
if $\psi$ is positive or of compact support.

The construction of $\theta$ states thus naturally requires,
see eqs.(5.1) and (3.6),
\lq\lq non--local boundary conditions\rq\rq, i.e. an ergodic mean
over boundary variables, integrated with
almost periodic functions $\psi(q) = \exp i \theta q \, R(q)$,
$R(q)$  periodic.

\medskip
For finite time intervals, the functional measure
$d\gamma^\theta_T $
on the circle,
eq. (5.1), has an interpretation  in
terms of winding numbers  of the trajectories
$\varphi(t) \in S^1 / 2\pi $, $-T \leq t \leq T $.
{}From eq.(5.1) one obtains in fact
$$ d\gamma^\theta_T
    = \sum_{n= -\infty}^\infty e^{i n \theta} \,
    d\gamma_{n,T}  \, e^{2 E_\theta T}    \eqno(5.3)   $$
where $d\gamma_{n,T}$ is the restriction to the trajectories
with winding number  $n$ of the positive
measure defined by the kernel
$   \sum_{m \in \interi}  K_t (\vp_1 , \vp_2 + 2 \pi m)$
(the factor $\exp 2 E_\theta T $ is necessary for the normalization
 $\int d\gamma^\theta_T = 1$).
The restriction to finite volume is crucial:
the mean square winding number
$< n^2 > \  \equiv \  \sum_n n^2 \int  d\gamma_{n,T}
  \; / \, \sum_n \int  d\gamma_{n,T}  $
goes to infinity as $T \nearrow \infty$,
and a winding number representation never exists in infinite volume.
Moreover, since
$ \sum_n   \int d\gamma_{n,T} \;  e^{2 E_\theta T} \
\sim \   e^{2 E_\theta T} $,
the sum over the winding numbers in eq.(5.3) is never done with
a probability measure, for $\theta \neq 0$.

For $\theta = 0$, eq.(5.3) do define a probability
measure over $n \in \interi$; the
positive functional defined by such a measure on
the algebra $\A_n$ generated by $\exp i \alpha n$,
$\alpha \in [0, 2\pi)$, converges, for $T \nearrow \infty$,
to the functional given by the ergodic mean over $n$.
A (generalized) probabilistic interpretation,
(as a measure on the Gelfand spectrum of $\A_n$)
is therefore allowed, {\it only for\/} $\theta = 0$,
for the winding number in infinite volume,
as a variable \lq\lq uniformly distributed over the
integers\rq\rq.

\vfill\eject

\immediate\closeout\fileref
                \par\vfill\eject
                \null\msk
                \centerline{\bf References}
                \bsk
                \input ref.tmp

\bye